\documentclass[12pt,preprint]{aastex}

\usepackage{epsf}

\usepackage{graphicx}

\begin{document}

\title{Magnetohydrodynamic Effects in Propagating Relativistic Jets: Reverse Shock and Magnetic Acceleration}

\author{
Yosuke Mizuno\altaffilmark{1,2}, Bing Zhang\altaffilmark{2}, Bruno Giacomazzo\altaffilmark{3}, Ken-Ichi Nishikawa\altaffilmark{1}, Philip E. Hardee\altaffilmark{4}, Shigehiro Nagataki\altaffilmark{5}, Dieter H. Hartmann\altaffilmark{6} }

\altaffiltext{1}{Center for Space Plasma and Aeronomic Research, University of Alabama in Huntsville, NSSTC, 320 Sparkman Drive, Huntsville, AL 35805, USA; mizuno@cspar.uah.edu.} 

\altaffiltext{2}{Department of Physics and Astronomy, University of Nevada, Las Vegas, NV 89154, USA.} 

\altaffiltext{3}{Max-Plank-Institut f\"{u}r Gravitationsphysik, Albert-Einstein-Institut, Potsdam-Golm, Germany.}

\altaffiltext{4}{Department of Physics and Astronomy, University of Alabama, Tuscaloosa, AL 35487, USA.} 

\altaffiltext{5}{Yukawa Institute for Theoretical Physics, Kyoto Univeristy, Sakyo, Kyoto, Japan.}

\altaffiltext{6}{Department of Physics and Astronomy, Clemson University, Clemson, SC 29634, USA.}

\shorttitle{MHD Effects in Propagating Relativistic Jets}


\begin{abstract}
We solve the Riemann problem for the deceleration of an arbitrarily
magnetized relativistic flow injected into a static unmagnetized
medium in one dimension. We find that for the same initial Lorentz 
factor, the reverse shock becomes progressively weaker with increasing
magnetization $\sigma$ (the Poynting-to-kinetic energy flux ratio),
and the shock becomes a rarefaction wave when $\sigma$ exceeds a
critical value, $\sigma_c$, defined by the balance between the magnetic
pressure in the flow and the thermal pressure in the forward shock.
In the rarefaction wave regime, we find that the rarefied region 
is accelerated to a Lorentz factor that is significantly larger than 
the initial value. This acceleration mechanism is due to the strong 
magnetic pressure in the flow. We discuss the implications of these 
results for models of gamma-ray bursts and active galactic nuclei.
\end{abstract}


\section{Introduction}

Relativistic jets are believed to exist in active galactic nuclei
(AGNs), black hole binaries, and gamma-ray bursts (GRBs), but their
composition is still poorly understood. It has been argued that magnetic 
fields could play an important dynamic role in these jets
(e.g. Lovelace 1976; Blandford 1976; Blandford \& Znajek 1977;
Blandford \& Payne 1982; Usov 1992; Thompson 1994; M\'esz\'aros \&
Rees 1997; Lyutikov \& Blandford 2003; Vlahakis \& K\"onigl 2003, 2004),
but the degree of magnetization, quantified by the
magnetization parameter $\sigma$ (the ratio of electromagnetic
to kinetic energy flux), is poorly constrained by observations. 
GRB afterglow modeling indicates that the ejecta are more magnetized 
than the ambient medium, suggesting a possibly important dynamic role for 
magnetic fields in GRB jets (Fan et al. 2002; Zhang, Kobayashi \& 
M\'esz\'aros 2003; Kumar \& Panaitescu 2003; Gomboc et al. 2008).

A useful diagnostic for the degree of jet-magnetization can be obtained 
from the interaction between the decelerating jet and the ambient
medium. Added magnetic field pressure in the jet alters the
condition for formation of a reverse shock (RS) as well as its
strength (Kennel \& Coroniti 1984). Analytical studies of the
deceleration of a GRB fireball with arbitrary magnetization (Zhang
\& Kobayashi 2005, hereafter ZK05; see also Fan et al.\ 2004 for 
$\sigma \leq 1$) suggest novel behavior that does not exist in pure 
hydrodynamic (HD) ($\sigma=0$) models (Sari \& Piran 1995; Kobayashi et al.\ 1999). However, consensus on the conditions required for the existence of the 
RS or how Poynting flux is transferred to kinetic flux in the interaction 
region has not yet been achieved (ZK05; Lyutikov 2006; Giannios et al. 2008).
We present a one-dimensional (1-D) study of the interaction between a 
magnetized relativistic flow and a static, unmagnetized external medium. 
A Riemann problem is solved both analytically and numerically over a broad 
range of $\sigma$.

\section{The Riemann Problem}

We consider a Riemann problem consisting of two uniform initial states
(left and right) with discontinuous hydrodynamic properties specified
by the rest-mass density $\rho$, gas pressure $p$, specific internal
energy $u$, specific enthalpy $h \equiv 1 + u/\rho c^{2} + p/\rho
c^{2}$, and normal velocity $v^{N}$. The right state (the medium external 
to the jet) is assumed to be a cold fluid with constant density, at rest. 
Specifically, we select the initial
conditions: $\rho_{R}=1.0 \rho_{0}$, $p_{R}=10^{-2} \rho_{0} c^2$,
$v^{N}_{R}=v^{x}_{R}=0.0$, where $\rho_{0}$ is an arbitrary
normalization constant (our simulations are scale-free) and $c$ is the
speed of light. The left state (the propagating relativistic
flow) is assumed to have a higher density and pressure
than the right state, as well as a relativistic velocity. Specifically, 
$\rho_{L}=10^{2} \rho_{0}$, $p_{L}=1.0 \rho_{0} c^2$, and $v^{N}_{L}=
v^{x}_{L}=0.995c$ ($\gamma_{L} \simeq 10$). The fluid is described by 
an adiabatic equation of state $p\propto \rho^{\Gamma}$ with $\Gamma=4/3$.

To investigate the effects of magnetic fields, we consider a perpendicular 
field component in the jet with $B^{y} =31.623$, $100.0$, $316.23$, and
$447.21$ in units of $(4\pi\rho_{0} c^{2})^{1/2}$ measured in the
laboratory frame, corresponding to
$\sigma \equiv B^{2}/4\pi\gamma^{2} h \rho c^2 \simeq
B^{2}/4\pi\gamma^{2} \rho c^2$ being $0.1$, $1.0$, $10.0$, and $20.0$,
respectively. This field is motivated by the predicted
toroidal field domination at the deceleration radius for GRB outflows
(e.g., Spruit et al.\ 2001; ZK05). Increasing $\sigma$ increases the total 
(kinetic plus magnetic) energy density of the left (jet) state.

\section{Results}

\begin{figure}
\epsscale{1.0}
\plotone{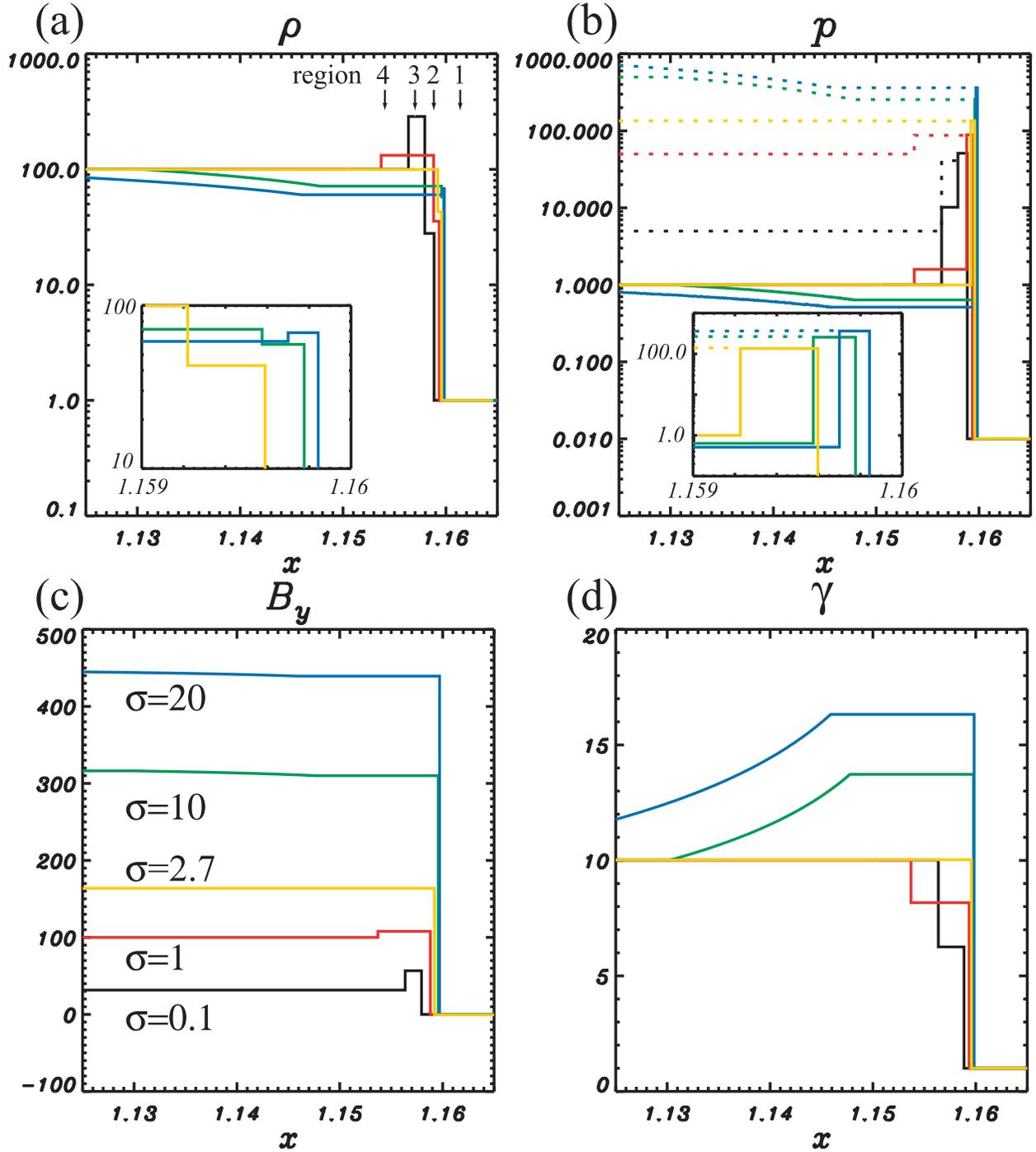} 
\caption{Profiles of (a) density, (b)  gas pressure ({\it solid 
lines}) and magnetic pressure ({\it dotted lines}), (c) magnetic 
field ($B^{y}$), and (d) Lorentz factor ($\gamma$) of $\sigma=0.1$ 
(black), $1.0$ (red), $10.0$ (green), $20.0$ (blue) cases at time 
$t=0.16$. Other parameters: $\rho_L=100.0$, $\rho_R=1.0$,
$\gamma_L=10.0$. The critical value, $\sigma_{c} \simeq 2.7$ case 
is shown as yellow lines. Close-up forward shock regions are inserted.
Arrows indicate four physically distinct regions: (1) unshocked medium, 
(2) shocked medium, (3) shocked flow and (4) unshocked flow corresponding
to the $\sigma=0.1$ case.
\label{f1}}
\end{figure}

We calculate exact solutions of this problem, using the code 
of Giacomazzo \& Rezzolla (2006), in the region $0.8 \le x \le 1.2$ with 
an initial discontinuity at $x=1.0$, where $x$ is in arbitrary units.

\subsection{Flow-Medium Interaction}

The exact solutions are presented in Figure 1. The four panels display
profiles of the gas density, gas (and magnetic) pressure, magnetic
field strength $B^y$, and the Lorentz factor at time
$t=0.16$\footnote{Here $t$ is in units of $x/c$ with $c=1$.}.
Different colors represent different $\sigma$ values: 0.1 (black), 1.0
(red), 2.7 (yellow), 10.0 (green) and 20.0 (blue). The initial Lorentz 
factor of the left state (jet) is $\gamma_L=10$.

For $\sigma=0.1$ (black), the solution shows a right-moving fast shock
(FS: forward shock; $S_{\rightarrow}$), a left-moving fast shock (RS: reverse
shock; $_{\leftarrow}S$) relative to the contact discontinuity ($C$). 
In the laboratory frame, the contact discontinuity and the two shocks 
move to the right.

For $\sigma=1.0$ (red), the solution shows similar profiles
($_{\leftarrow}SCS_{\rightarrow}$) as for $\sigma=0.1$. The FS is stronger 
(due to a higher jump in pressure) and slower (more deceleration relative to 
the frame of the contact discontinuity), while the RS is weaker but faster. 
These features are expected from analytical work (ZK05; Giannios et al.\ 
2008), and agree with 1-D relativistic MHD simulations (Mimica et al.\ 2007, 2008).

When the magnetization of the flow exceeds $\sigma=2.7$, the shock
profiles change drastically (the significance of this particular 
value of $\sigma$ is discussed below). For $\sigma=10.0$ (green)
and $\sigma=20.0$ (blue), a prominent left-going rarefaction wave
($_{\leftarrow}R$) is observed, instead of a left-going shock 
(see also Romero et al. 2005; Mimica et al. 2007). When the
rarefaction wave propagates into the jet flow, density and gas
pressure decrease, and the flow velocity increases. The terminal Lorentz 
factor of the left (jet) state and the FS region reaches $\gamma 
\sim 14$ for $\sigma=10$ and $\gamma > 16$ for $\sigma=20$. This 
{\em magnetic acceleration} mechanism stems from the magnetic 
pressure in the flow\footnote{We note that Romero et al.\ (2005) and
Mimica et al.\ (2007) also discovered the rarefaction wave regime
discussed in this paper, but did not investigate the magnetic acceleration 
mechanism and its astrophysical implications in detail.}.

This magnetic acceleration mechanism is solely a MHD effect and requires
the magnetic field to generate a rarefaction wave. This is 
different from the HD/MHD boost mechanism proposed by Aloy
\& Rezzolla (2006), and further investigated by Mizuno et al.\ (2008) and 
Aloy \& Mimica (2008). The HD/MHD mechanism is a purely relativistic 
mechanism, which invokes a relativistic flow perpendicular to the propagation 
direction of the rarefaction wave. The mechanism discussed here occurs even 
in the Newtonian case, and acts parallel to the propagation direction of the 
rarefaction wave. In general, the acceleration efficiency is smaller than that 
of the HD/MHD boost mechanism (see \S3.3 for more discussions).

\subsection{Conditions for Reverse Shock or Magnetic Acceleration}

The magnetic pressure profiles (dotted lines) in Fig. 1(b) reveal the
physical condition required for the transition from a reverse shock to
a rarefaction wave. It is evident that the magnetic pressure increases
as $\sigma$ increases. In the reverse shock cases ($\sigma=0.1,1$),
the upstream magnetic pressure is lower than the gas pressure in the
forward shock, while in the rarefaction wave cases ($\sigma=10,20$),
the upstream magnetic pressure exceeds the gas pressure in the FS. Thus, 
the balance between the upstream magnetic pressure in the 
unshocked flow region and the FS gas pressure in the shocked 
medium (ZK05; Romero et al. 2005) provides the condition separating the 
two regimes. This condition can be derived analytically (see also ZK05).
For the interaction between a relativistic flow and an
external medium, there exist four physically distinct regions: (1)
unshocked medium, (2) shocked medium, (3) shocked flow, and (4)
unshocked flow. Hereafter, $Q_{i}$ denotes the value of a quantity
$``Q''$ in region $``i''$. From the relativistic shock jump conditions 
with $\Gamma = 4/3$, one can write $u_{2}/\rho_{2}c^{2}=(\gamma_{2}-1) 
\simeq \gamma_{2}$ and $\rho_{2}/\rho_{1} = 4 \gamma_{2} +3 \simeq 
4\gamma_{2}$. 
A constant speed across the contact discontinuity requires
$\gamma_2=\gamma_3$, and the relation between the gas pressure and 
the internal energy gives $p_2 = u_2/3$. 
Thus, the thermal pressure generated in the FS region is 
$p_2 =(1/3) (\gamma_2-1)(4\gamma_2+3) \rho_1 c^2 \simeq (4/3)
\gamma_2^2 \rho_1 c^2$.
The pressure balance condition is $B_4^2/8\pi\gamma_4^2 \sim p_2$, because 
in pressure balance there is no reverse shock or rarefaction wave, so that 
region 4 and region 3 are matched as $B_4 \simeq B_3$, and $\gamma_4 \simeq 
\gamma_3=\gamma_2$. Using the definition $\sigma \equiv \sigma_4 = B_4^2/4\pi
\gamma_4^2 \rho_4 c^2$, one can derive a critical $\sigma_c$ value
\begin{equation}
\sigma_c = \frac{2}{3}(\gamma_4-1)(4\gamma_4+3) \frac{\rho_1}{\rho_4}
\simeq \frac{8}{3}\gamma_4^2 \frac{\rho_1}{\rho_4}=
\frac{8}{3}\gamma_L^2 \frac{\rho_R}{\rho_L}~.
\label{sigmac}
\end{equation}
The condition for the existence of a reverse shock is $\sigma < \sigma_c$, 
which is Eq.\ (31) of ZK05. The condition for a rarefaction
wave and magnetic acceleration is $\sigma > \sigma_c$.
We adopted $\rho_{1}=\rho_{R}=1.0$,
$\rho_{4}=\rho_{L}=10^{2}$, and $\gamma_{4}=\gamma_{L}=10.0$, so that the 
critical value is $\sigma_c \simeq 2.7$. Our calculations indicate that
$\sigma_c$ marks the transition point where neither a reverse shock nor a
rarefaction wave is established (yellow lines in Fig. 1). To verify this 
for a larger parameter space, we
investigate the $\sigma$-dependences of various quantities in detail.
Fig. 2(a) shows the gas pressure in 
the region through which the reverse shock/rarefaction wave has
propagated. Initial Lorentz factors are $\gamma_{L} = 5$,
$10$, and $20$, respectively. For all cases, we fix the
flow density at $\rho_{L} = 10^{2}$ and increase $B$ (hence $\sigma$). 
The total initial energy density of the flow increases with $\sigma$.
In all cases, the gas pressure decreases with
$\sigma$ smoothly without a sharp transition from the RS
regime (solid lines) to the reverse 
rarefaction wave regime (dotted lines). The critical
magnetization parameters are $\sigma_c \simeq 0.7, 2.7, 10.6$ 
for $\gamma_{L}=5, 10, 20$, respectively, derived from
the analytical solution Eq.(\ref{sigmac}). We notice that in
the RS regime, the strength of the shock decreases
rapidly with increasing $\sigma$. The critical magnetization 
parameter $\sigma_c$ increases with $\gamma_{L}$, so that a RS 
can exist in the high-$\sigma$ regime if $\gamma_L$ is sufficiently 
large (see also ZK05).

Another commonly invoked RS-condition states that the shock speed in the
fluid frame (region 4) is higher than the speed of the Alfv\'{e}n wave, i.e.
$\gamma'_{\rm RS,RR} > \gamma'_A$, where $\gamma'_{\rm RS,RR}=(\gamma_4
/\gamma_{\rm RR,RS}+\gamma_{\rm RR,RS}/\gamma_4)/2$, $\gamma'_A
=(1+\sigma)^{1/2}$, and $\gamma_{\rm RS,RR}$ is the Lorentz factor
of the reverse shock or reverse rarefaction wave in the laboratory frame, 
which can be calculated using the exact
solution of Giacomazzo \& Rezzolla (2006). Giannios et al.
(2008) claim that this condition is different from the pressure 
balance condition. However, in Figures 2(b) and 2(c), we present the ratios 
of gas pressure in the FS region to magnetic pressure in the flow, 
$p_{FS}/p_{B}$, and $\gamma'_{\rm RS,RR}/\gamma'_A$,
and find that both ratios reach unity at the same critical value 
$\sigma_c$. This suggests that the two reverse shock conditions
have the same physical origin, at least for the 1-D model studied here.

\begin{figure}
\epsscale{1.0}
\plotone{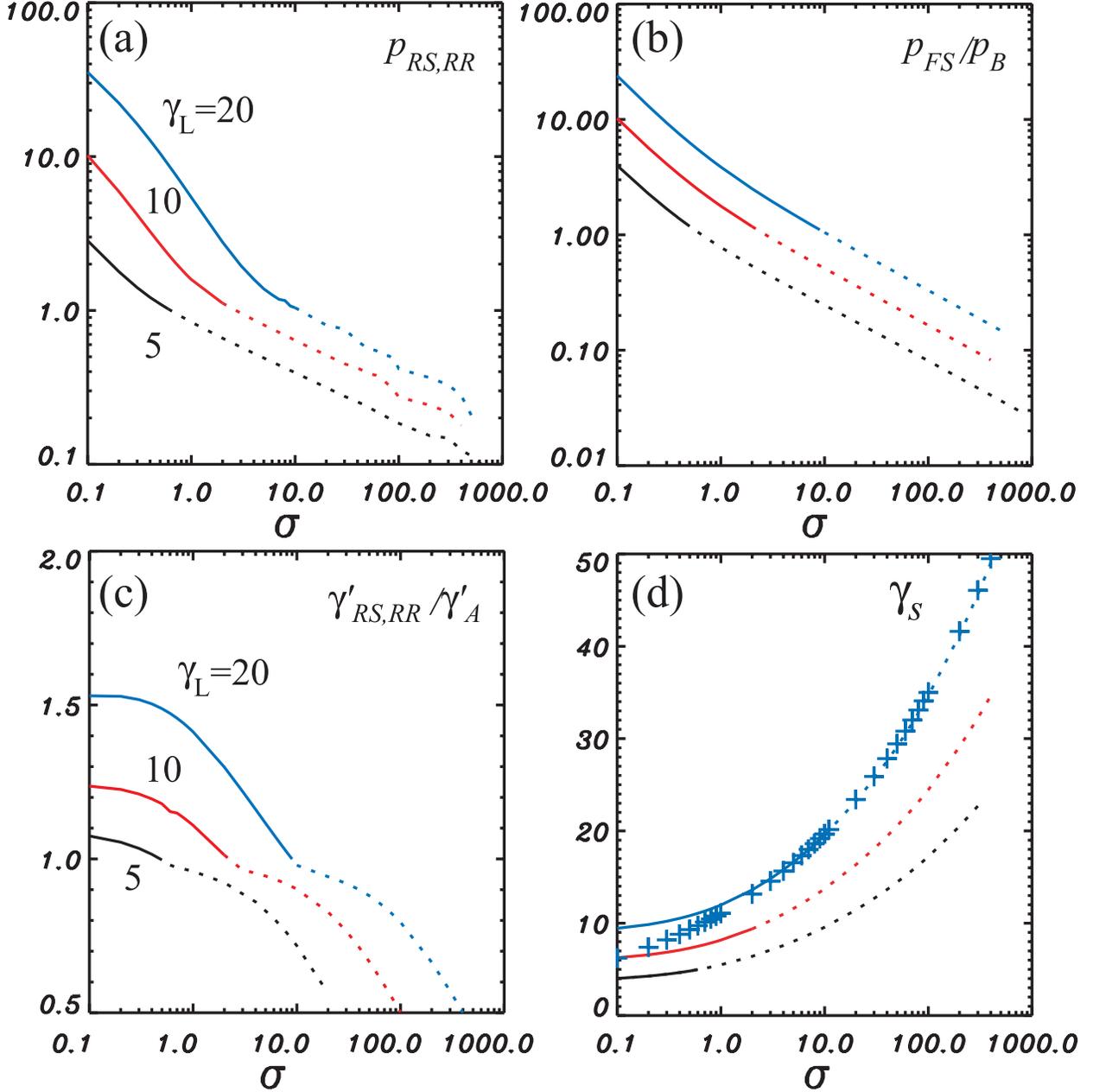} 
\caption{The $\sigma$-dependences of (a) gas pressure in the region through
which the reverse shock (RS; solid lines)/rarefaction wave 
(RR; dotted lines) has propagated; (b) the ratio of the gas pressure 
in the forward shock region to the magnetic pressure in the flow;
(c) the ratio of the Lorentz factor of the propagating 
reverse shock/rarefaction wave to the Alfv\'{e}n Lorentz factor 
in the rest frame of the fluid; (d) the maximum Lorentz factor
in the shocked region, in the exact solution. Different initial
Lorentz factors have been calculated: 
$\gamma_{L}=5$ (black), $10$ (red), and $20$ (blue). Crosses are 
the values of the estimated terminal Lorentz factor in the 
$\gamma_{L}=20$ case according to Eq.(2). A constant flow density 
is adopted, so that the total initial
energy density of the flow increases with $\sigma$.
\label{f2}}
\end{figure}

\subsection{Terminal Lorentz Factor and Magnetic Acceleration Efficiency}

To better understand the magnetic acceleration mechanism, we 
plot the Lorentz factor as a function of $\sigma$ in Fig. 2(d). 
For the magnetic acceleration case,
this is the terminal Lorentz factor after acceleration. 
Because of the dependence of $\sigma_c$ on $\gamma_L$,
a higher $\sigma$ is needed to achieve acceleration for a higher 
$\gamma_L$. The terminal Lorentz factor can be estimated analytically
by requiring that the thermal pressure in the FS region balances the 
magnetic pressure in the region through which the rarefaction wave has 
propagated. For the terminal Lorentz factor 
$\gamma_t$, this condition can be expressed (roughly) as 
$B_3^2/8\pi\gamma_t^2 = (1/3) (\gamma_t-1)(4\gamma_t+3) 
\rho_1 c^2 \simeq (4/3) \gamma_t^2 \rho_1 c^2$.
From the definition of $\sigma$ with $B_{4} \simeq B_{3}$, this becomes
\begin{equation}
\gamma_{t} \simeq 
\left( { 3 \gamma_{4}^{2} \sigma \rho_{4} \over 8 \rho_{1}}
\right)^{1/4} 
\label{gammat}.
\end{equation}
Crosses in Fig.2(d) denote values of terminal Lorentz factors
calculated from Eq.(\ref{gammat}) for model parameters, 
$\gamma_{4}= \gamma_{L} = 20$, 
$\rho_{1}=\rho_{R}=1.0$, and $\rho_{4} = \rho_{L} = 10^{2}$,
in good agreement with the exact solution of the Riemann 
problem in the reverse rarefaction wave regime.

\begin{figure}[t]
\epsscale{1.0} 
\plotone{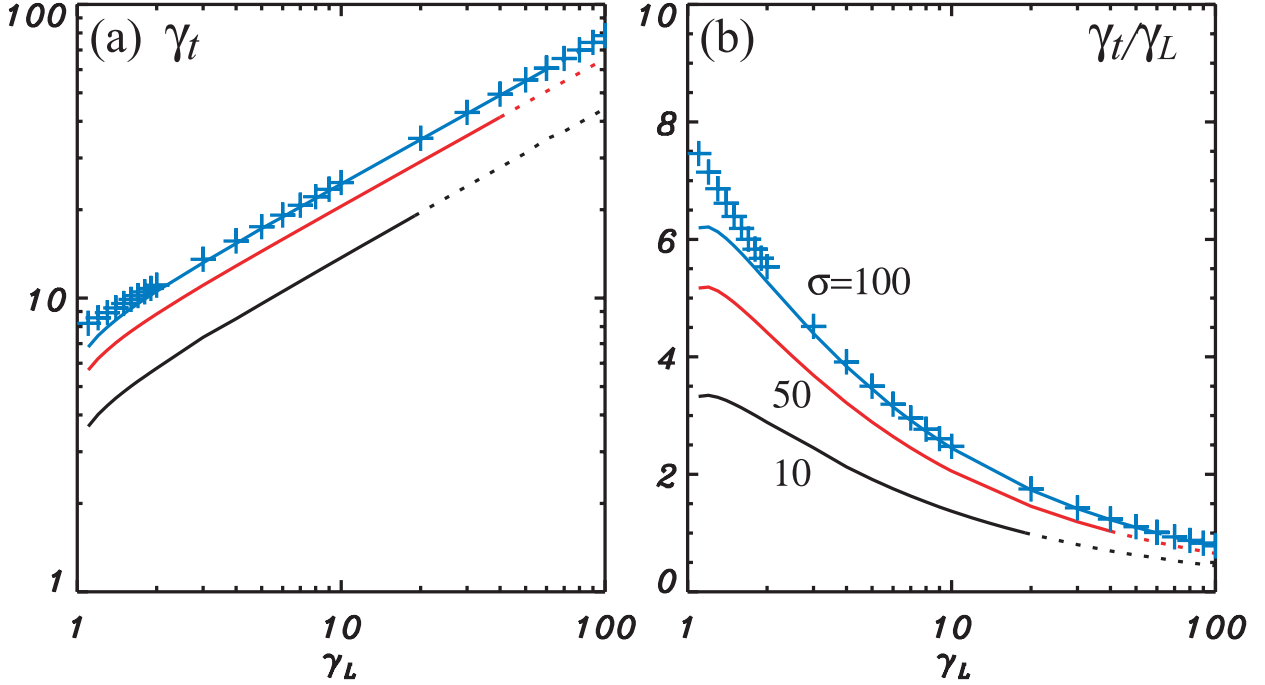} 
\caption{The dependences of (a) the terminal Lorentz factor $\gamma_t$;
and (2) the acceleration efficiency $\gamma_t/\gamma_L$ on the initial
Lorentz factor $\gamma_L$. Solid lines are for the RR regime and 
dotted lines are for the RS regime. 
Different inital magnetizations have been calculated: $\sigma=10$ (black), 
$50$ (red), and $100$ (blue). Crosses are values estimated with Eq.(2) 
for $\sigma=100$. \label{f3}}
\end{figure}

To investigate the acceleration efficiency, we present in Fig. 3 the terminal
Lorentz factor $\gamma_t$, and its ratio to the initial Lorentz factor 
($\gamma_{t}/\gamma_{L}$) as a function of the initial flow Lorentz factor 
$\gamma_L$. While a flow with a higher initial Lorentz factor reaches a higher 
terminal Lorentz factor, a lower initial Lorentz factor implies a higher 
acceleration efficiency. From Eq.(\ref{gammat}) it follows that 
$\gamma_{t}/\gamma_{4} \simeq (3 \sigma \rho_{4} / 8 \rho_{1})^{1/4} 
\gamma_{4}^{-1/2}$, in good agreement with the exact solution 
of the Riemann problem in relativistic regime.

\section{Summary and Discussion}

We solved the 1-D Riemann problem for the deceleration of an
arbitrarily magnetized relativistic flow in a static, unmagnetized
medium. For the same initial Lorentz factor, the
reverse shock becomes progressively weaker with increasing
$\sigma$, turning into a rarefaction wave when $\sigma \ge \sigma_c$, at 
which point the magnetic pressure in the
flow is balanced by the thermal pressure in the forward shock. In the
rarefaction wave regime, material in the FS region is accelerated 
due to the strong magnetic pressure in the flow. This magnetic
acceleration mechanism may thus play an important role in the dynamics
of strongly magnetized, relativistic flows.

Numerical MHD simulations (e.g.\ Koide et al.\ 1999, 2000; Nishikawa
et al.\ 2005; Mizuno et al.\ 2007) are essential to understand
magnetized relativistic jets. We performed 1-D special
relativistic MHD simulations of a relativistic flow propagating in an
external medium using the RAISHIN code (Mizuno et al.\ 2006). The
simulation results are in good agreement with the exact solution (Giacomazzo \&
Rezzolla 2006), serving as a test of the RAISHIN code. As MHD simulations can 
tackle problems for which an exact solution is not known, we plan to utilize 
RAISHIN to solve more realistic configurations (e.g., relativistic shells 
with a finite width and conical geometry, as often envisaged in the GRB problem).

The magnetic acceleration mechanism discussed here also applies in
the Newtonian MHD limit. The transition point from a reverse shock to 
a rarefaction wave is then also given by the pressure balance condition,
and the terminal velocity of the flow can be estimated from the Newtonian 
shock jump condition as
$v_{t}= c_{s1} ( p_2/p_1 - 1 ) \sqrt{2/\Gamma/[(\Gamma+1)(p_2/p_1)
+(\Gamma-1)]}$, where $c_{s1}=(\Gamma p_1/\rho_1)^{1/2}$ is the
sound speed in the upstream medium (see also Hawley et al. 1984). 
Defining $\sigma = (B_4^2/8\pi) / (\rho_4 v_4^2/2)$ in the Newtonian limit, 
one can derive the terminal velocity of the flow, $v_{t}$ determined by balance 
between the magnetic pressure and the pressure in the forward shock region, $p_{2} = 
B_{3}^{2}/ 8 \pi \simeq B_{4}^{2} / 8 \pi = \sigma \rho_{4} v_4^2/2$.
For a strong shock, ($p_2 \gg p_1$), the terminal velocity can be approximated as 
$v_{t} \simeq \sqrt{ \sigma (\rho_{4}/\rho_1)/ (\Gamma +1)} v_4$. 
For $v_t=v_4$, one derives $\sigma_c \simeq (\Gamma+1)\rho_1/\rho_4$. 
For $\Gamma=5/3$, typical for non-relativistic shocks, this expression is 
consistent with Eq.(1) for $\gamma_L=1$. Although the general physics is the same, 
the dependence of the terminal velocity $(v_t/c)\gamma_t$ is different 
in the Newtonian and relativistic case. 

Our results have implications for understanding deceleration of 
strongly magnetized outflows, possibly present in GRBs and AGNs. 
Exact solutions indicate that the condition for the existence of a 
reverse shock is $\sigma <\sigma_c$, as suggested by ZK05 (cf.\ Giannios 
et al.\ 2008). The paucity of bright optical flashes in GRBs 
(e.g., Roming et al. 2006) may, among other interpretions, be attributed 
to highly magnetized flows. Furthermore, the magnetic acceleration mechanism 
discussed here suggests that $\sigma$ and $\gamma$ are not
independent parameters at the deceleration radius. For high-$\sigma$
flows, the ejecta would experience magnetic acceleration at small radii,
before reaching the coasting regime, so that the coasting Lorentz
factor (i.e., the ``initial'' Lorentz factor for the afterglow) is at
least the ``terminal'' Lorentz factor defined by Eq.(\ref{gammat}). As
a result, the high-$\sigma$ and low-$\gamma$ part of parameter space is 
suppressed. This implies that some region in the $\xi-\sigma$ parameter 
space\footnote{$\xi$ is defined as $\sqrt{r_{dec}/r_{s}}$ (Sari
\& Piran 1995), where $r_{dec}$ is the deceleration radius where the ejecta accumulate 
from the external medium a mass $\gamma_{0}^{-1}$ times their own mass, and $r_{s}$ is 
the spreading radius where the width of the ejected shell starts to increase due to 
propagation of a sound wave.} of GRB models (ZK05; Giannios et al.\ 2008) is suppressed as well. 
Here we only focus on 1-D models with Cartesian geometry. 
Implications for GRB models will be discussed in more detail when this Riemann problem is solved in conical jet geometry.

Variable emission (down to minute timescales) observed in some TeV
blazars suggests very high Lorentz factors in AGN jets (Aharonian et
al.\ 2007). Models for the production of TeV emission appeal to
high Lorentz factor jet cores surrounded by lower Lorentz factor
sheaths (e.g., Ghisellini et al.\ 2005) or rapid jet deceleration
(Georganopoulos et al.\ 2005) in order to reconcile the required jet
Lorentz factors with lower Lorentz factors suggested by proper motion
studies. Our results suggests the possibility of magnetic
acceleration occuring where highly magnetized jet material overtakes
more weakly magnetized jet material. 
In this case the magnetically
accelerated Lorentz factor behind the forward shock can significantly
exceed the Lorentz factor of the overtaken jet material.

In this study we considerd a static, unmagnetized medium, but for 
a weakly magnetized medium, our conclusions hold as well. 
However, the conclusions will not apply in the very large $\sigma$ regime 
when the underlying MHD approximation breaks down, e.g., $\sigma \sim$ several 
100s in the problem of GRBs (Spruit et al.\ 2001; Zhang \& M\'esz\'aros 2002).

\acknowledgments
We thank S. Kobayashi, H. Sol, and L. Rezzolla for helpful comments.
Y.M. and B.Z. acknowledge NASA NNG05GB67G, NNG05GB68G, and NNX08AE57A 
for partial support during Y.M.'s stay at UNLV.
Y.M., P.H., and K.I.N. acknowledge partial support by NSF AST-0506719,
AST-0506666, NASA NNG05GK73G, NNX07AJ88G, and NNX08AG83G. 
B.G. thanks CSPAR-UAH/NSSTC for hospitality during the preparation of 
part of this work and acknowledges the DFG grant SFB/Transregio 7 for 
partial support. S.N. is partially supported by Grants-in Aid for 
Scientific Research of the Japanese Ministry of Education, Culture, 
Sports, Science and Technology 19047004, 19104006 and 19740139. The
simulations were performed on Columbia Supercomputer at NAS Division at
NASA Ames Research Center and Altix3700 BX2 at YITP in Kyoto University.

\end{document}